\documentclass[journal]{IEEEtran}
\usepackage{graphicx}
\usepackage{booktabs}
\usepackage{tabularx}
\usepackage{amsmath}
\usepackage{amssymb}

\ifCLASSINFOpdf
\else
\fi
\begin{document}
%
\title{Development of a 256-channel Time-of-flight\\Electronics System For Neutron Beam Profiling}
%
%
%

\author{Haolei~Chen,
        Changqing~Feng,
        Jiadong~Hu,
        Laifu~Luo,
        Li~Wang,
        Zhixin~Tan 
        and~Shubin~Liu
\thanks{C. Feng is the corresponding author, email : fengcq@ustc.edu.cn}
\thanks{H. Chen, C. Feng, J. Hu, L. Luo, L. Wang and S. Liu are from State Key Laboratory of Particle Detection and Electronics, Hefei 230026, China}
\thanks{H. Chen, C. Feng, J. Hu, L. Luo, L. Wang and S. Liu are from Department of Modern Physics, University of Science and Technology of China, Hefei 230026, China}
\thanks{Z. Tan is from Institude of High Energy Physics, Chinese Academy of Sciences (CAS), Beijing 100049, China}
\thanks{Z. Tan is from Dongguan Neutron Science Center, Dongguan 523803, China}
}

%
%

\markboth{Journal of \LaTeX\ Class Files,~Vol.~14, No.~8, August~2015}%
{Shell \MakeLowercase{\textit{et al.}}: Bare Demo of IEEEtran.cls for IEEE Journals}
%



\maketitle

\begin{abstract}
A 256-channel time-of-flight (TOF) electronics system has been developed for a beam line facility called ``Back-n WNS'' in China Spallation Neutron Source (CSNS). This paper shows the structure and performance of electronics system and the test results in CSNS.

A 256-channel photomultiplier tube (PMT) is chosen as the detector in this system. In order to acquire the time information from the PMT, an electronics system has been designed. The electronics system mainly includes one front-end board (FEB), four time-to-digital converter (TDC) boards and one clock distribution module (CDM). There are 256 channels on FEB and 64 channels on each TDC board. The FEB is connected to the PMT with high-density connectors and the TDC boards are connected to the FEB through 2m cables. The TDC boards are 6U size so that they can be PCI extensions for Instrumentation (PXI) cards. Data from TDC boards can be transferred to the PXI control card through the backboard. In order to make four TDC boards work synchronously, a CDM outputs four clock signals to TDC boards which are distributed from one clock source. The TDC boards achieve a timing resolution of 3.5ns by test with a signal generator. The TOF measurement system has been used in CSNS.

\end{abstract}

\begin{IEEEkeywords}
Time measurement, analog processing circuits, field programmable gate arrays
\end{IEEEkeywords}

%
\IEEEpeerreviewmaketitle

\section{Introduction}
\begin{figure}
\includegraphics[width=3.5in,clip,keepaspectratio]{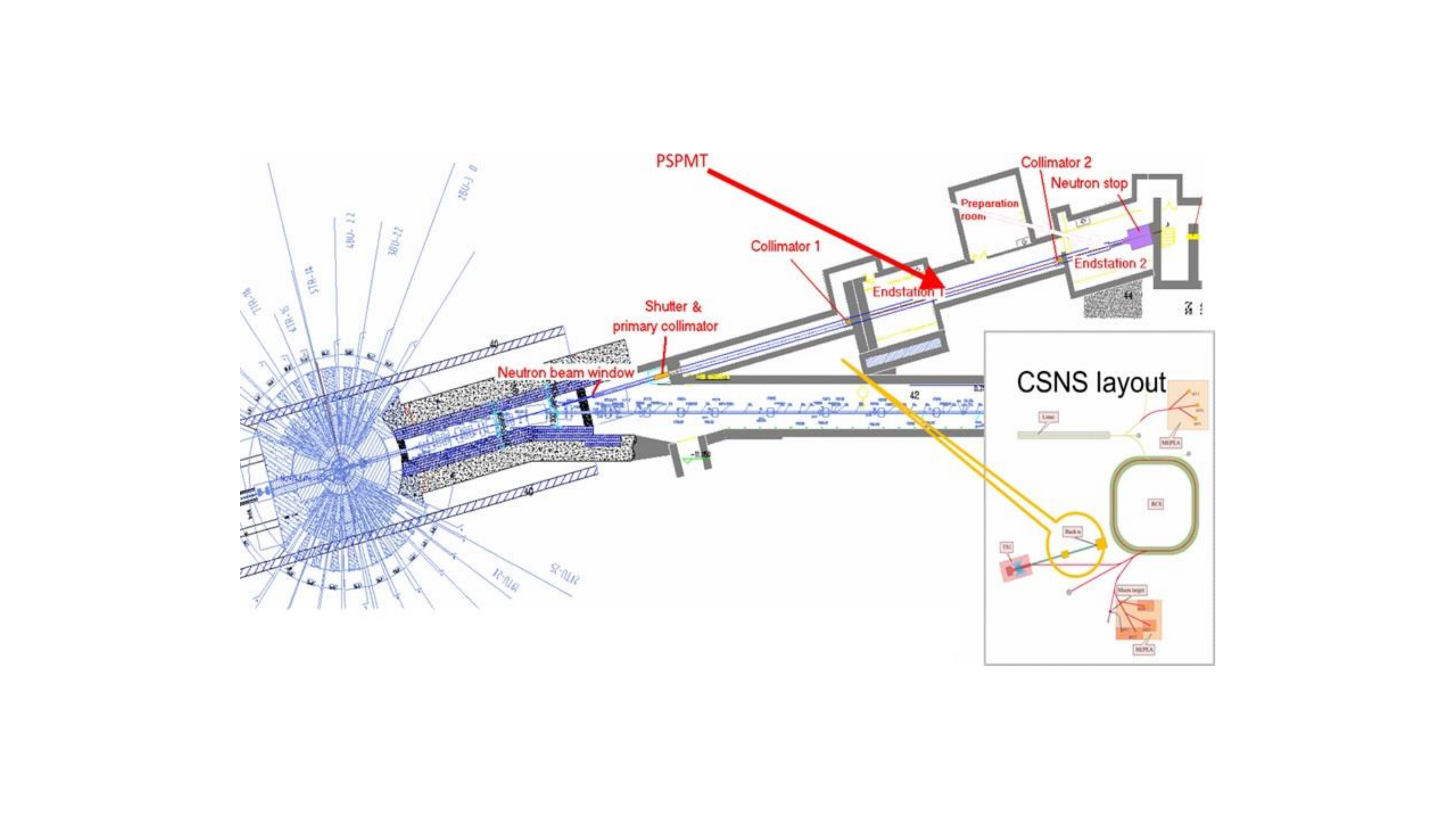}
\caption{Structure of Back-n WNS, the figure is from [1]}
\label{Fig.1}
\end{figure}

%
%
%
%
\IEEEPARstart{T}{he} CSNS is a large scientific device that generates neutrons by hitting the target with high-energy protons. The proton beam impinges the target at 25Hz and produces high flux neutrons [1]. A facility called ``Back-n WNS'' in CSNS exploits the application of back-streaming neutrons which have a very wide energy spectrum from eV to hundreds of MeV [2][3][4]. The structure of the facility is shown in Fig.~\ref{Fig.1}

\begin{figure}
\includegraphics[width=3.5in,clip,keepaspectratio]{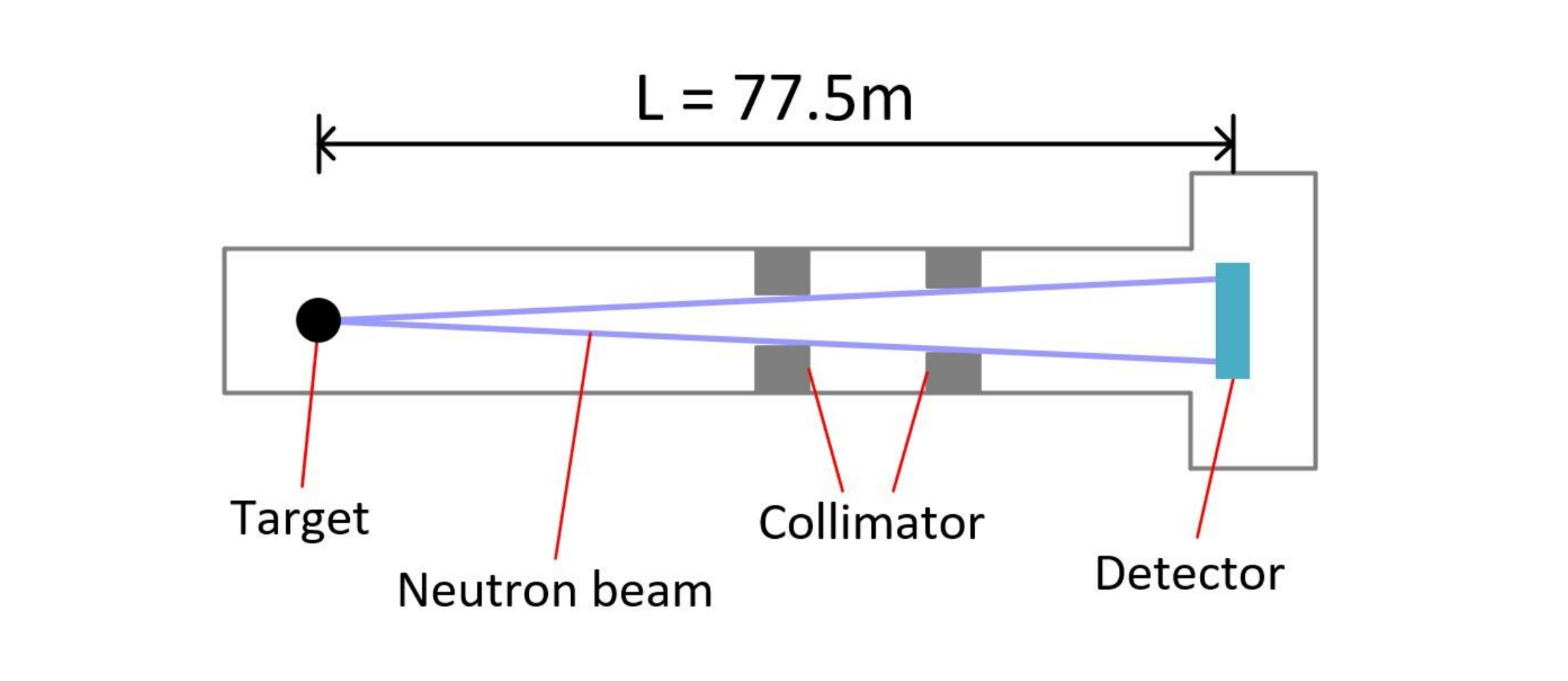}
\caption{Structure of TOF measurement system}
\label{Fig.2}
\end{figure}

Due to the neutron beam in Back-n beam line has high flux and wide energy spectrum, we can use it easily to complete neutron resonance radiography. For this idea, if we want to have a good image, the energy of neutrons should be measured to get the transmission information of them [1]. A proper method to acquire the energy is to measure the flight time when neutrons travel from the target to the detector. Also, a multi-channel detector is necessary for drawing an image. Thus, we designed a 256-channel TOF measurement system for the neutron resonance radiography. The structure of TOF measurement system is shown in Fig.~\ref{Fig.2}

For the speed of neutrons which have the energy from eV to hundreds of MeV is far less than the speed of light, there is a relation between the energy of neutron E, the quality of neutron m, the travelling distance L and the flight time t :$$E = \cfrac{1}{2} m (\cfrac{L}{t})^2$$In this experiment, the energy of neutrons we are focus on is from 0.5MeV to 10MeV for that they are the majority of the beam line and have not been touched by other neutron spallation source [1]. In order to make the energy resolution better than 1\% when the energy of neutron is 10MeV, we should control the time measurement precision better than about 9ns when the travelling distance L is 77.5m.

There are already some TOF measurement systems running now. The TOF system in BES III has 448 channels’ readout electronics. It uses an application specific integrated circuits (ASIC) called high-performance time-to-digital converter (HPTDC) to measure the time. The readout system achieved a timing resolution of 25ps [5]. The TOF system in AMS-02 also uses HPTDC to measure the time. In this structure, there is a low threshold and a high threshold to improve the measurement precision [6].

The HPTDC is designed by an electronics group at CREN. The chip can be used as 32 channels with a low resolution mode of 265ps time resolution [7].

Except for the use of HPTDC, there is also a method to measure time by field programmable gate array (FPGA). The advantage of this method is that we can easily change some parameters to match the need for the experiment and the cost of design can be lower.

In this system, due to the situation that the necessary precision is not very high and it's better to make the design more flexible, we used a FPGA to measure the time.

\section{Time-of-flight Measurement System}

Similar to the structures of experiments in BES III and AMS-02, our electronics system are mainly made up by two parts: the front-end electronics and the back-end electronics. The front-end electronics is used to be a discriminator which is to transform analog signals from the detector into digital signals with standard level. The back-end electronics takes the digital signals from front-end electronics for time measurement and transmit results to a storage device.

\begin{figure}
\includegraphics[width=3.5in,clip,keepaspectratio]{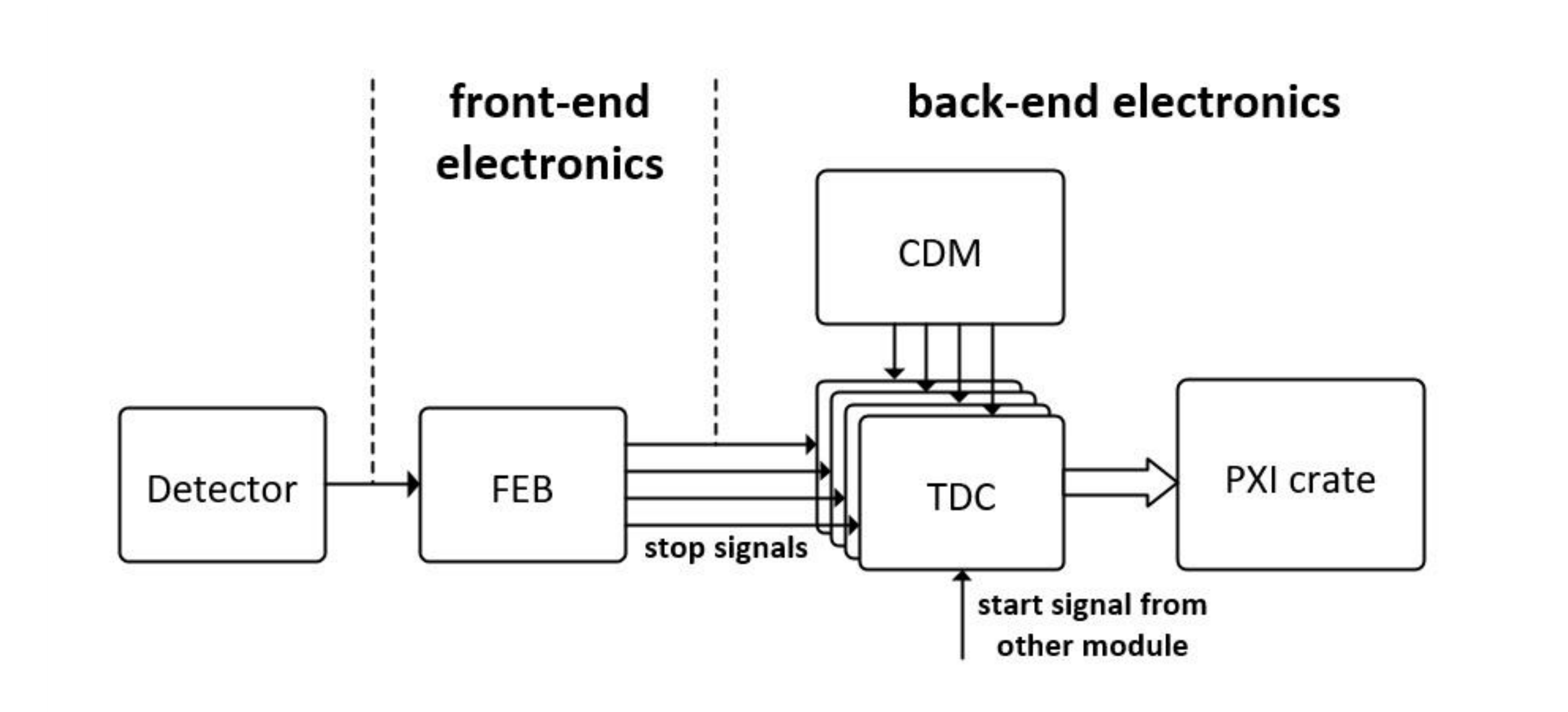}
\caption{Structure of TOF electronics system}
\label{Fig.3}
\end{figure}

In this system, the front-end electronics is made up by FEB and the back-end electronics consists of TDC, CDM and PXI crate. Fig.~\ref{Fig.3} shows the structure of TOF electronics system.

\begin{figure}
\includegraphics[width=3.5in,clip,keepaspectratio]{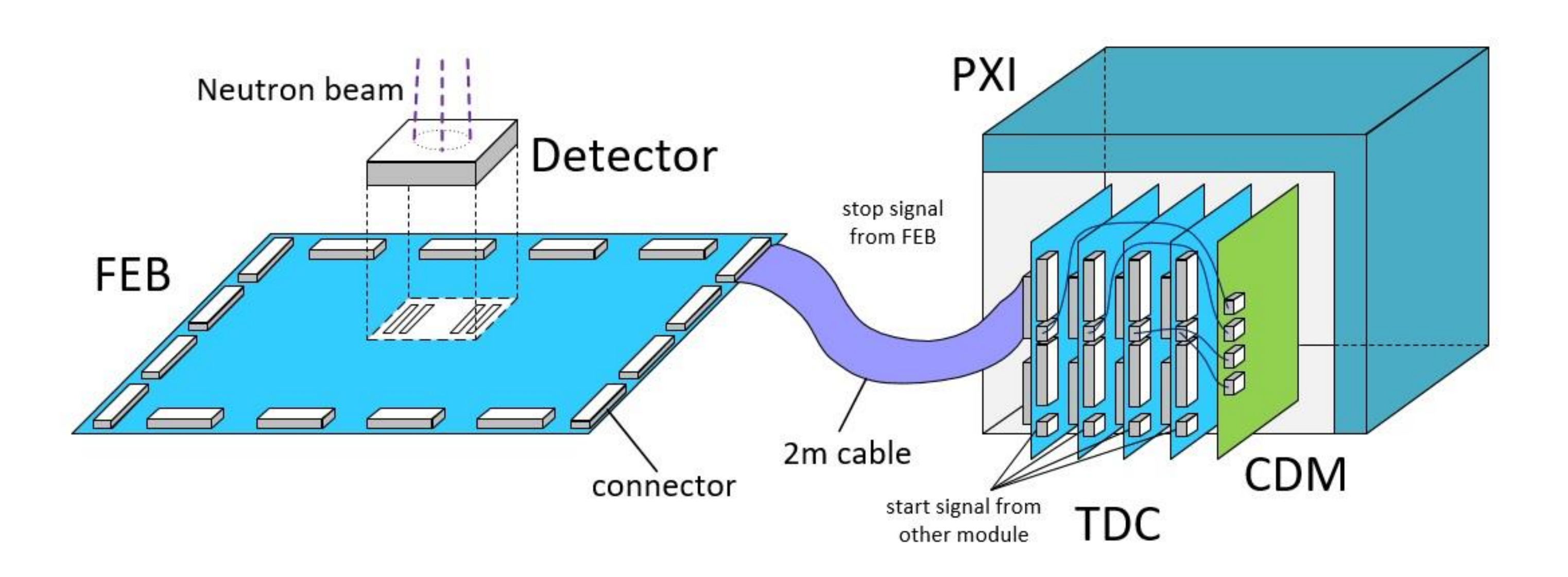}
\caption{Connection schematic for measurement system}
\label{Fig.4}
\end{figure}

The detector can be put on the FEB so they are on the beam line together for detection of neutrons. The back-end electronics is away from the beam line so the signals from FEB are transferred to the TDC boards through cables. The TDC boards and CDM are all housed in one PXI crate. The schematic of connection for the measurement system is shown in Fig.~\ref{Fig.4}

\subsection{Detector}

There are several materials like some scintillators which can detect neutrons for that they can emit light when neutrons hit them. Then a PMT receives the light and changes it into photoelectrons. The photoelectrons multiply in the PMT and turn into current signals. So, a detector composed by scintillator and PMT can be used to recognize neutrons and generate a current signal. Due to the time of emitting light by scintillator and multiplication of electrons in PMT is short, we can consider that the leading edge of the current signal stands the time when the particle hits the detector.

\begin{figure}
\includegraphics[width=3.5in,clip,keepaspectratio]{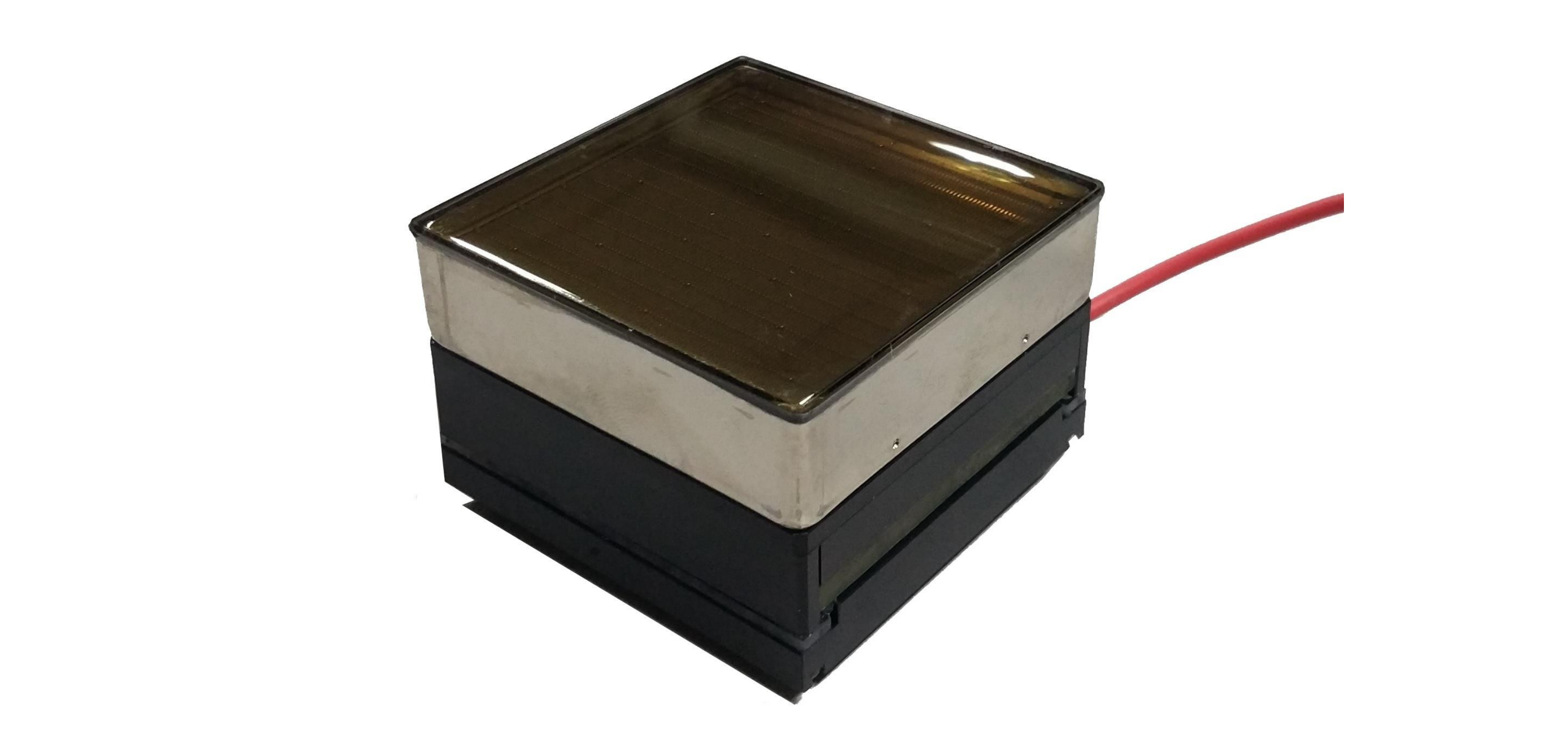}
\caption{The appearance of H9500}
\label{Fig.5}
\end{figure}

We use a material including $^6$Li as the scintillators and H9500 as the PMT which is designed by Hamamatsu. The rise time of the leading edge of the signal by H9500 is about 0.8ns [8]. There are 256 channels in one PMT and the effective area is 49mm$\times$49mm. The gain of PMT is up to 10$^6$ when the supply voltage is 950V. Fig.~\ref{Fig.5} shows the appearance of H9500.

\subsection{FEB}

Because there are not suitable ASIC for the discriminator circuit, we are going to use discrete components to process the 256-channel signals from the detector.

Due to the situation that the outputs of the detector are analog current signals, we need to convert them into voltage signals for further process. In addition, since the signals from the detector usually have small amplitude and short pulse width, they should be amplified and compared to thresholds and turn to standard amplitude pulse signals. So, we use amplifiers and comparators to process the signals from the detector.

In order to amplify the signals with about 0.8ns leading edge from the detector, the bandwidth of the amplifier should be high enough. Since the system has to deal with 256 channel signals, the amplifier and comparator should also be as more channels as possible.

So we choose LMH6722 designed by Texas Instruments as the amplifier with four channels in one chip and AD8564 designed by Analog Devices as the comparator with four channels in one chip to process the signals. The LMH6722 has a bandwidth of 400MHz when the output is 500mVpp and the gain is 2V/V [9]. The amplifiers and the comparators are powered by +5V and -5V. Since the signal from the detector is a negative pulse, we set the threshold of the comparators from -5V to 0V which can be adjusted by adjustable resistances so that we can select the signals which we are interested in.

The signals from the comparator are TTL standard signals, we convert them to differential LVDS standard signals to transfer to TDC boards through cables so the signals have stronger ability of anti-interference.

\begin{figure}
\includegraphics[width=3.5in,clip,keepaspectratio]{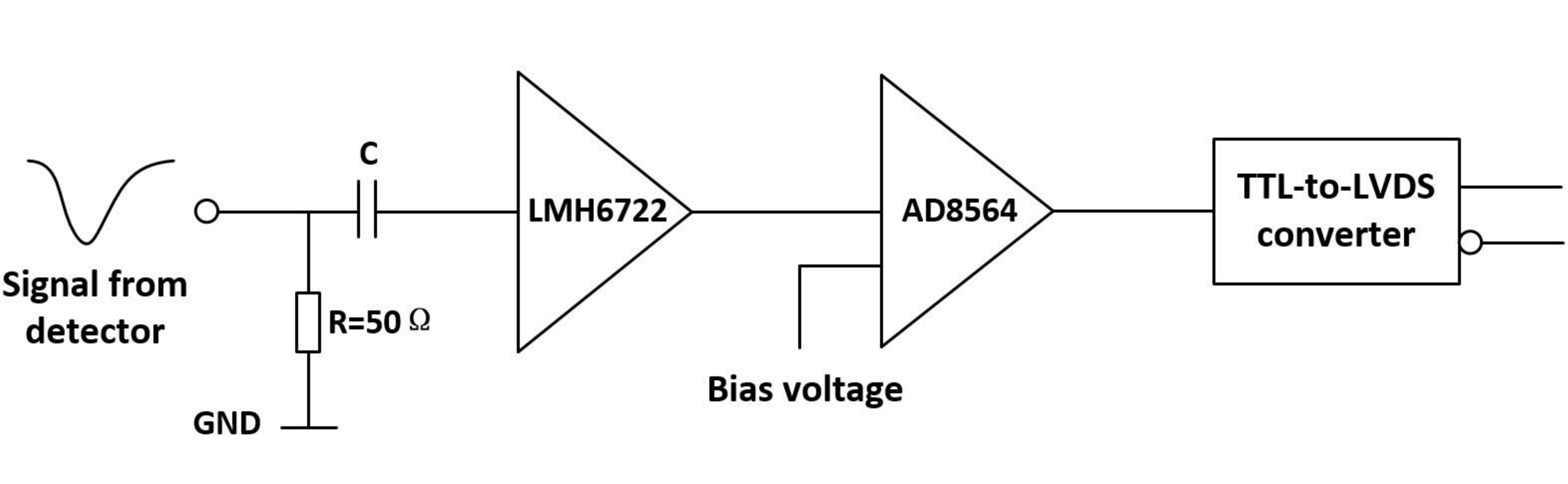}
\caption{The structure of FEB}
\label{Fig.6}
\end{figure}

\begin{figure}
\includegraphics[width=3.5in,clip,keepaspectratio]{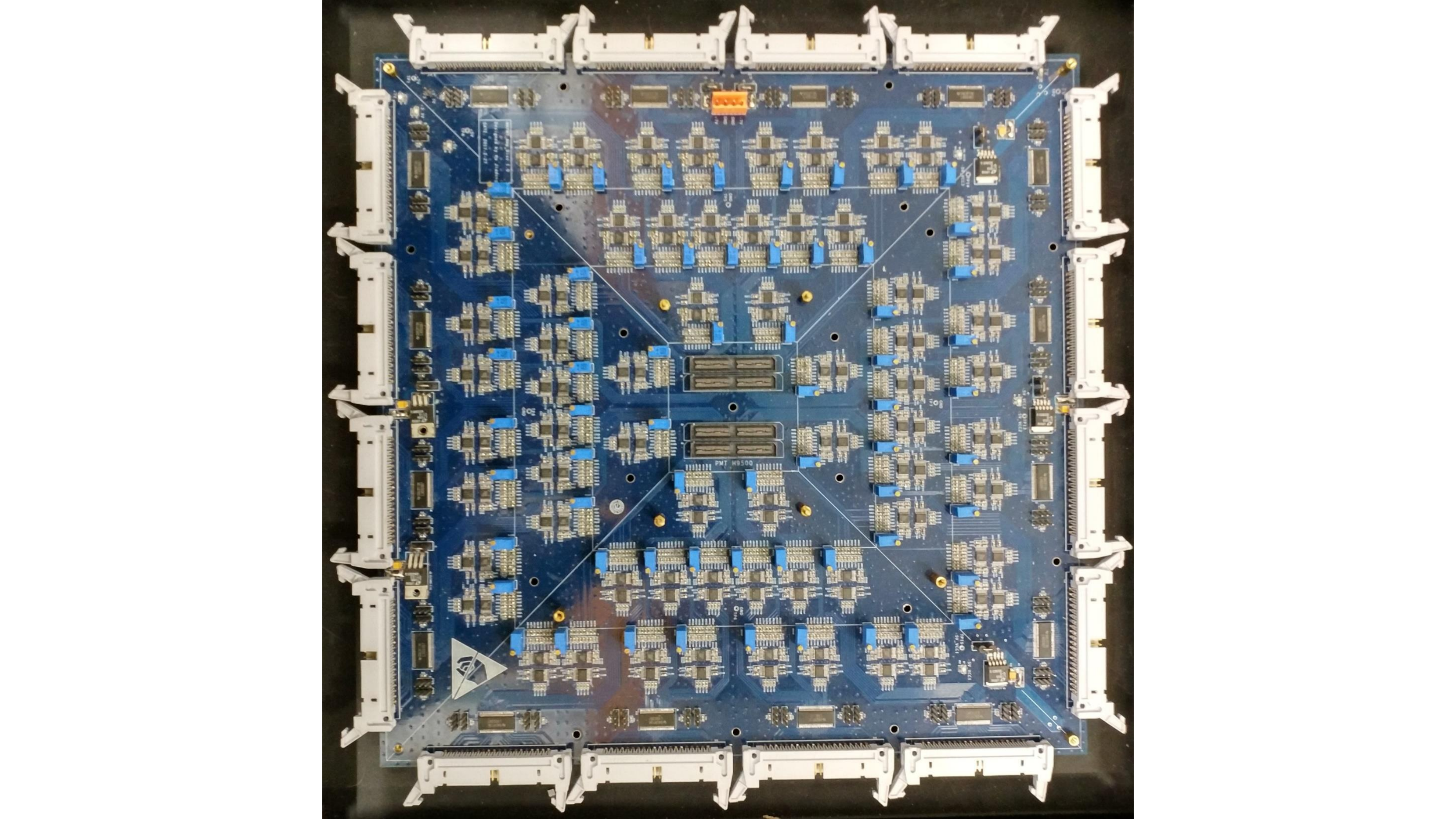}
\caption{The PCB of FEB}
\label{Fig.7}
\end{figure}

Fig.~\ref{Fig.6} shows the structure of FEB and Fig.~\ref{Fig.7} shows the printed circuit board (PCB) of FEB.

\subsection{TDC board}

TDC board, as the name implies, is used to measure time and output a data which can stand for how long the time is. When the proton beam begins to fly to the target, the system gives out a pulse signal. The TDC boards take this signal as the start signal. When the neutrons hit the detector, the FEB outputs pulse signals to the TDC boards as the stop signals. Then TDC boards measure the time between the leading edge of start signal and stop signals in FPGAs.

\begin{figure}
\includegraphics[width=3.5in,clip,keepaspectratio]{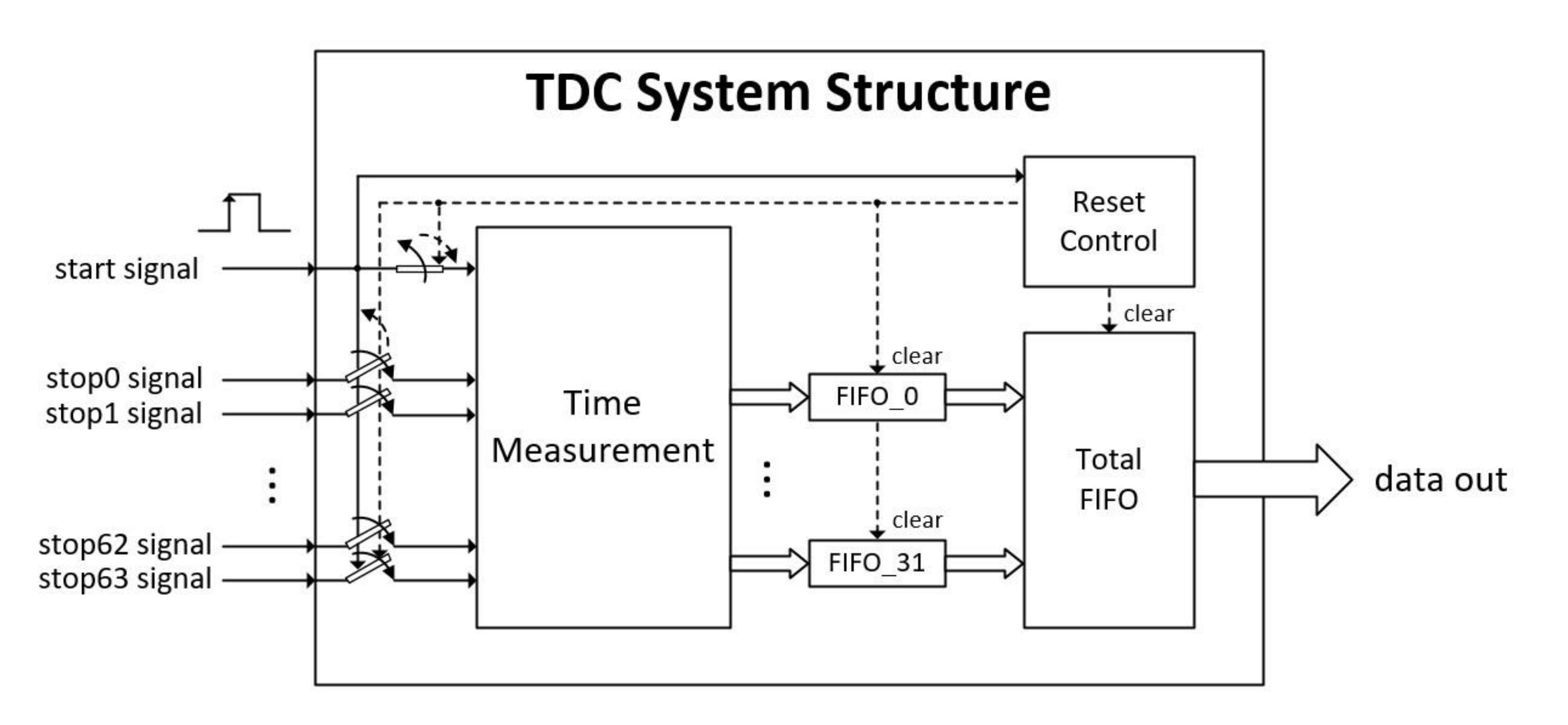}
\caption{The structure of TDC system in FPGA}
\label{Fig.8}
\end{figure}

The TDC system in FPGA is driven by the start signal. When the system detects the start signal, it begins to receive stop signals and prevent new start signals. In this design, because the flight time of neutrons which has energy of 1eV is about 5.6ms. So we make the stop signals which are 0 to 10ms after the start signal are valid for the system which is enough to collect the signals we are interested in. Every two stop channels’ data will be cached into a FIFO temporarily. When the moment comes which is 10ms after the start signal, the system prevents new stop signals going in. Due to the fact that the collision is at a frequency of 25Hz, there are still 15ms left. So we can take several milliseconds for remaining data transmission. After this time, the system will reset and wait for the next start signal. Fig.~\ref{Fig.8} shows the structure of TDC system in FPGA.

\begin{figure}
\includegraphics[width=3.5in,clip,keepaspectratio]{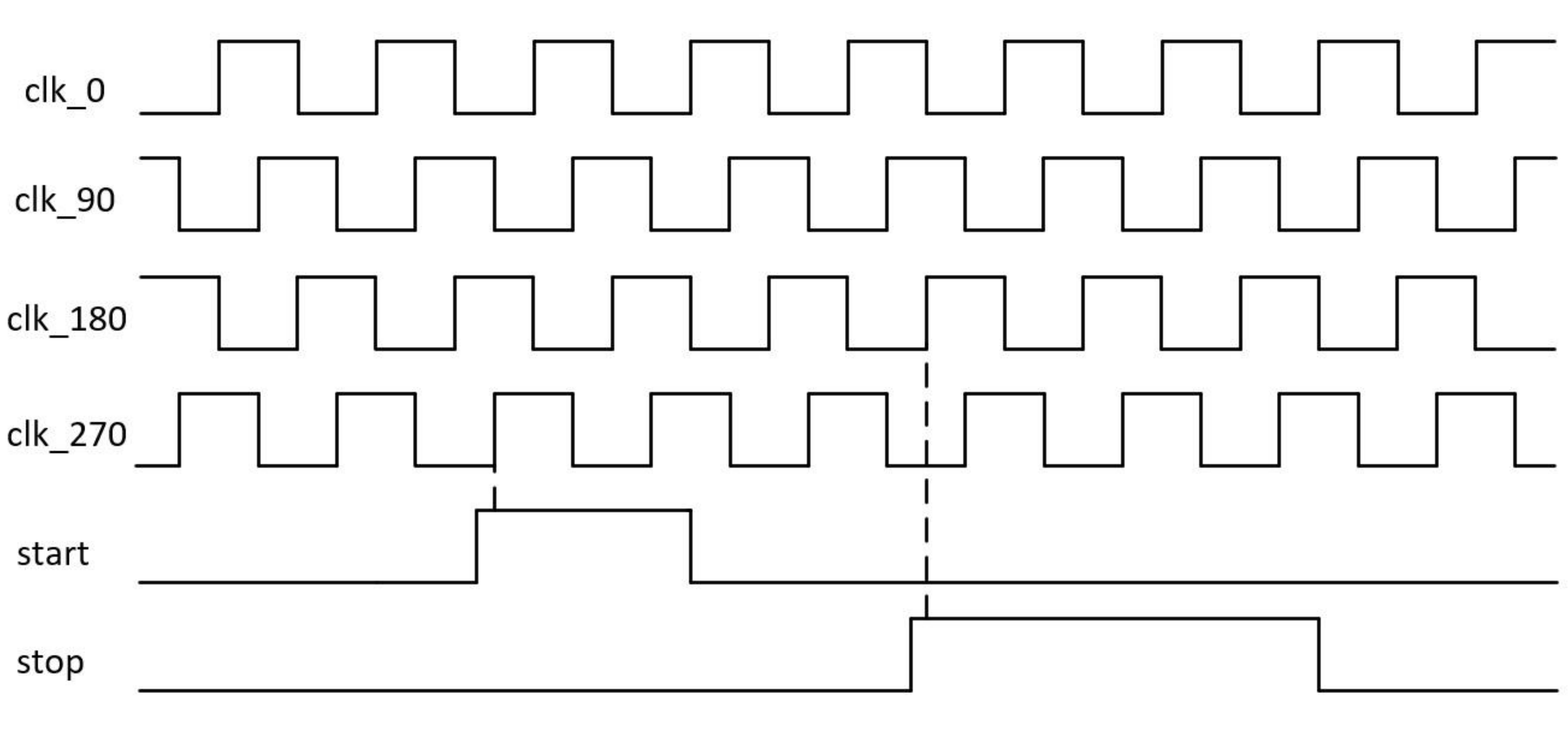}
\caption{Principle of clock phase separation technology}
\label{Fig.9}
\end{figure}

A 22-bits counter is used in FPGA to measure the time which is driven by a 40MHz clock. The dynamic range is from 0 to about 100ms. Because the precision of the counter depends on the period of the clock, we can only get 25ns time resolution which is not enough for our experiment. So we use a method called clock phase separation technology to improve the measurement accuracy. If we separate the clock signal into four signals, which have 0 phase delay, 90 phase delay, 180 phase delay and 270 phase delay compared to the input clock signal, we can get 6.25ns time resolution. Fig.~\ref{Fig.9} shows the principle of clock phase separation technology. When the hit signal (start signal or stop signal) arrives, the system in FPGA judges which leading edge of four clock signal will first arrive. Then different results will be coding into different data. Therefore, the time measurement system can achieve a 6.25ns resolution.

\begin{figure}
\includegraphics[width=3.5in,clip,keepaspectratio]{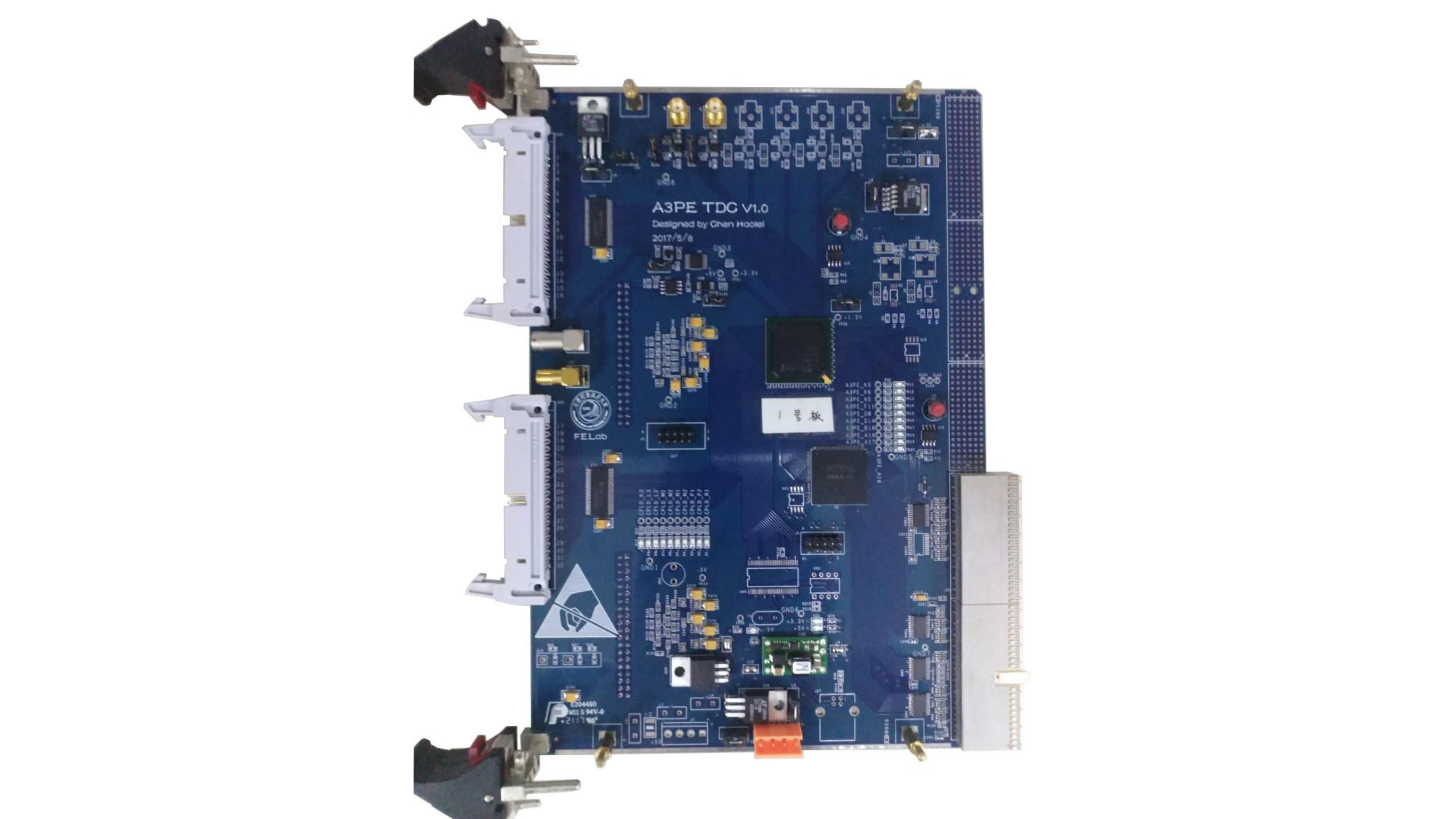}
\label{Fig.10}
\end{figure}

Each TDC board can deal with 64 channels’ signals from FEB. So the TOF system includes four TDC boards. Fig.~\ref{Fig.10} shows one of the PCB of TDC boards. The size of TDC boards is 6U which is fit for the PXI crate.

\subsection{CDM}

\begin{figure}
\includegraphics[width=3.5in,clip,keepaspectratio]{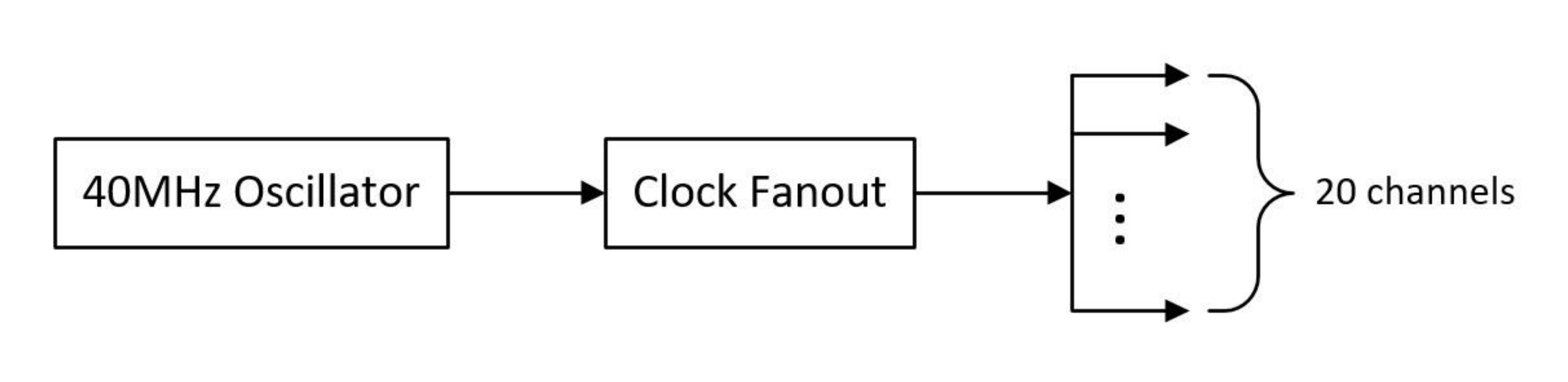}
\caption{The structure of CDM}
\label{Fig.11}
\end{figure}

A clock distribution module is used to support four clock signals to the TDC boards which are from the same clock source. The size of CDM is 6U so that it can also be PXI card. Fig.~\ref{Fig.11} shows the structure of CDM.

\subsection{PXI crate}

Because TDC boards can output data which stands the time measurement result, we need a method to transfer data to a storage device. PXI control card can be used to obtain data from TDC boards. The crate also can supply voltage for them.

\section{Test Result}

\subsection{TDC board performance test}

\begin{figure}
\includegraphics[width=3.5in,clip,keepaspectratio]{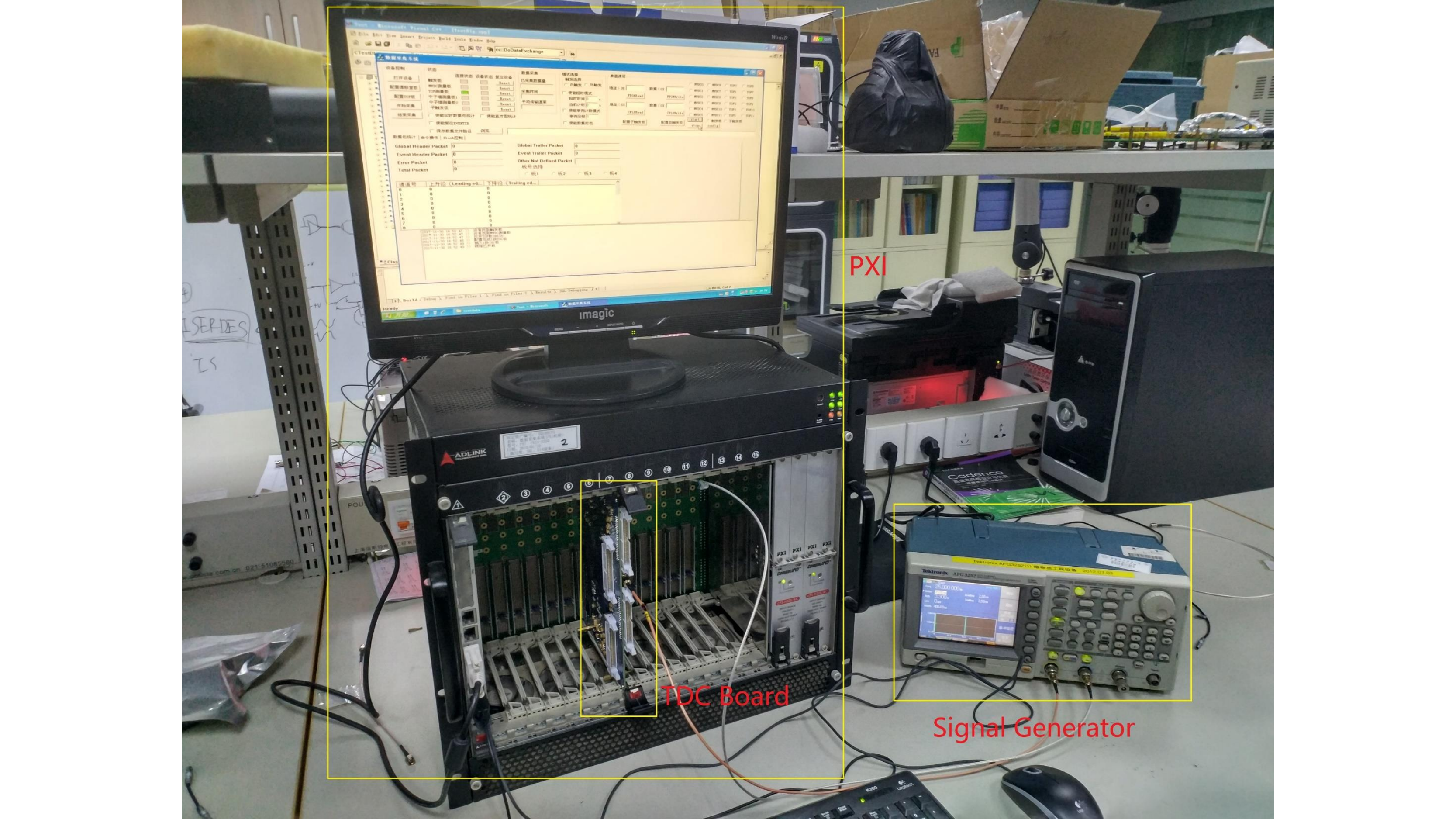}
\caption{Test platform of TDC board}
\label{Fig.12}
\end{figure}

We used a signal generator AFG3252 designed by Tektronix to test the time measurement precision of TDC board. The signal generator outputs two pulse signals, one as the start signal and the other as the stop signal, to the TDC board. We can adjust the time between the leading edge of start signal and stop signal as different input time. Fig.~\ref{Fig.12} shows the test platform of TDC board.

\begin{figure}
\includegraphics[width=3.5in,clip,keepaspectratio]{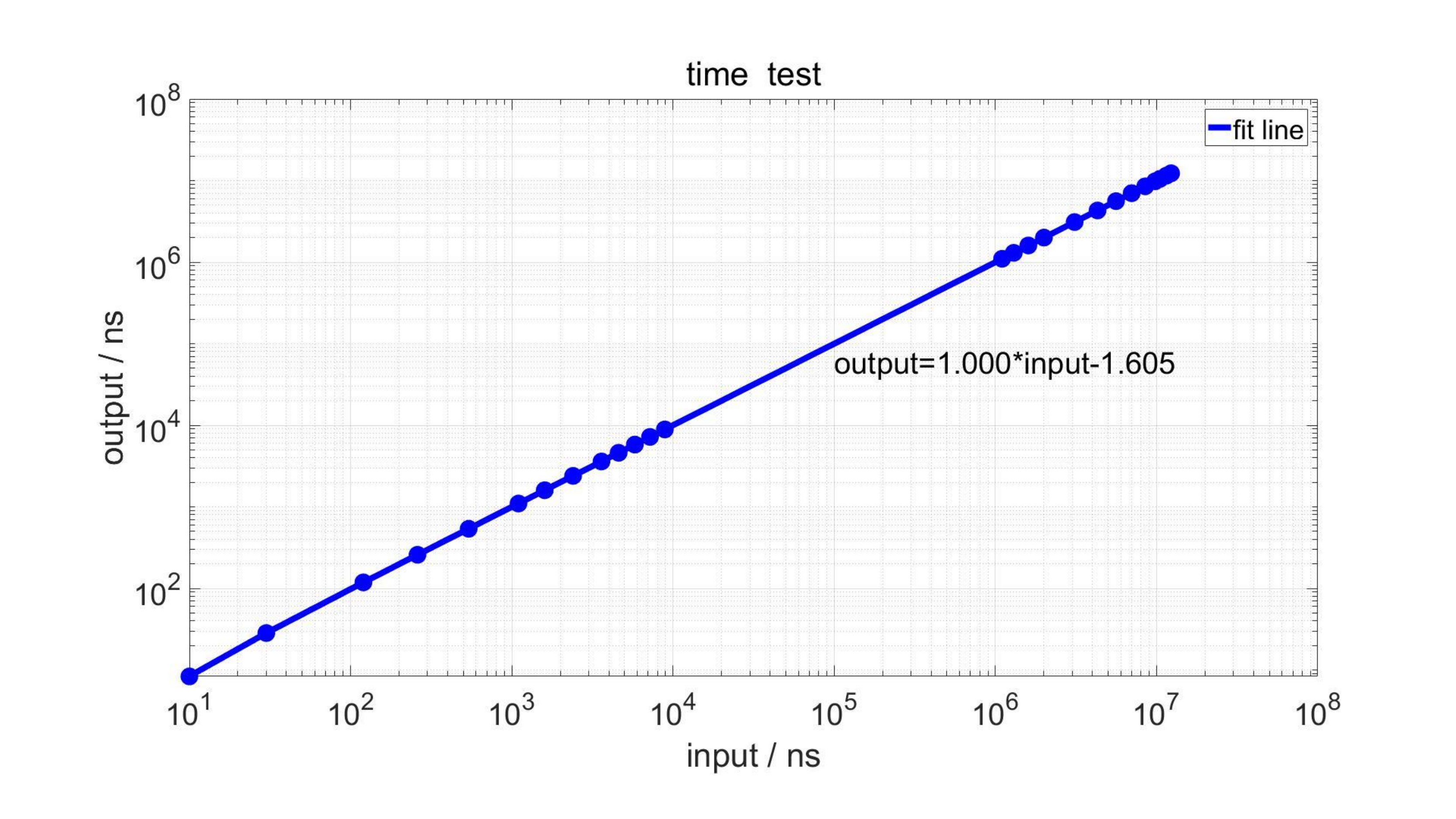}
\caption{Linearity test result of TDC board}
\label{Fig.13}
\end{figure}

\begin{figure}
\includegraphics[width=3.5in,clip,keepaspectratio]{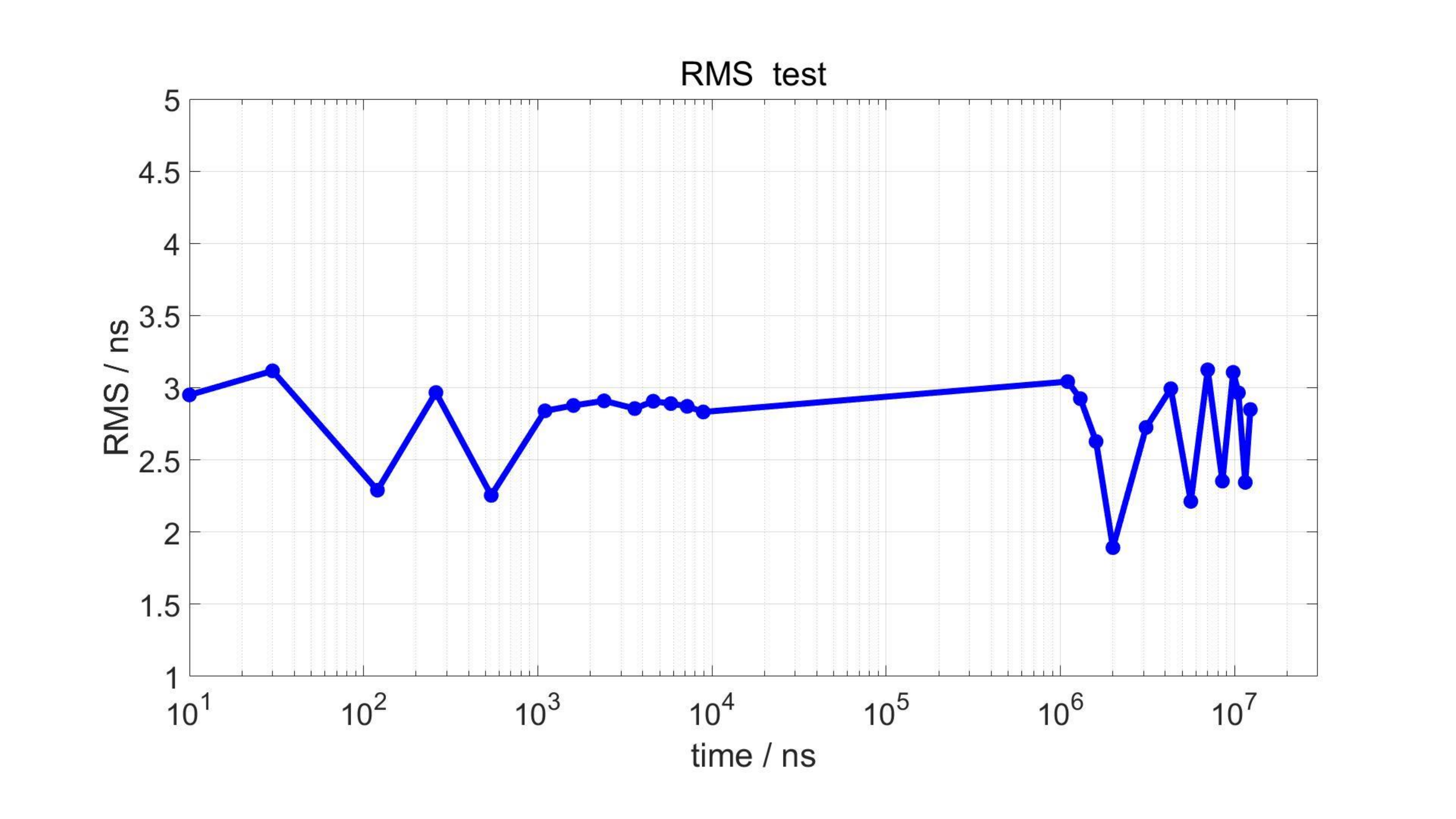}
\caption{Time precision test result of TDC board}
\label{Fig.14}
\end{figure}

Fig.~\ref{Fig.13} shows how the output changes by the input and Fig.~\ref{Fig.14} shows how the time measurement precision changes by the input.

The linearity test result shows that TDC board has a precise response among about 10ms dynamic range. The precision test shows that RMS is below 3.5ns among the dynamic range.

\subsection{TOF system test in CSNS}

\begin{figure}
\includegraphics[width=3.5in,clip,keepaspectratio]{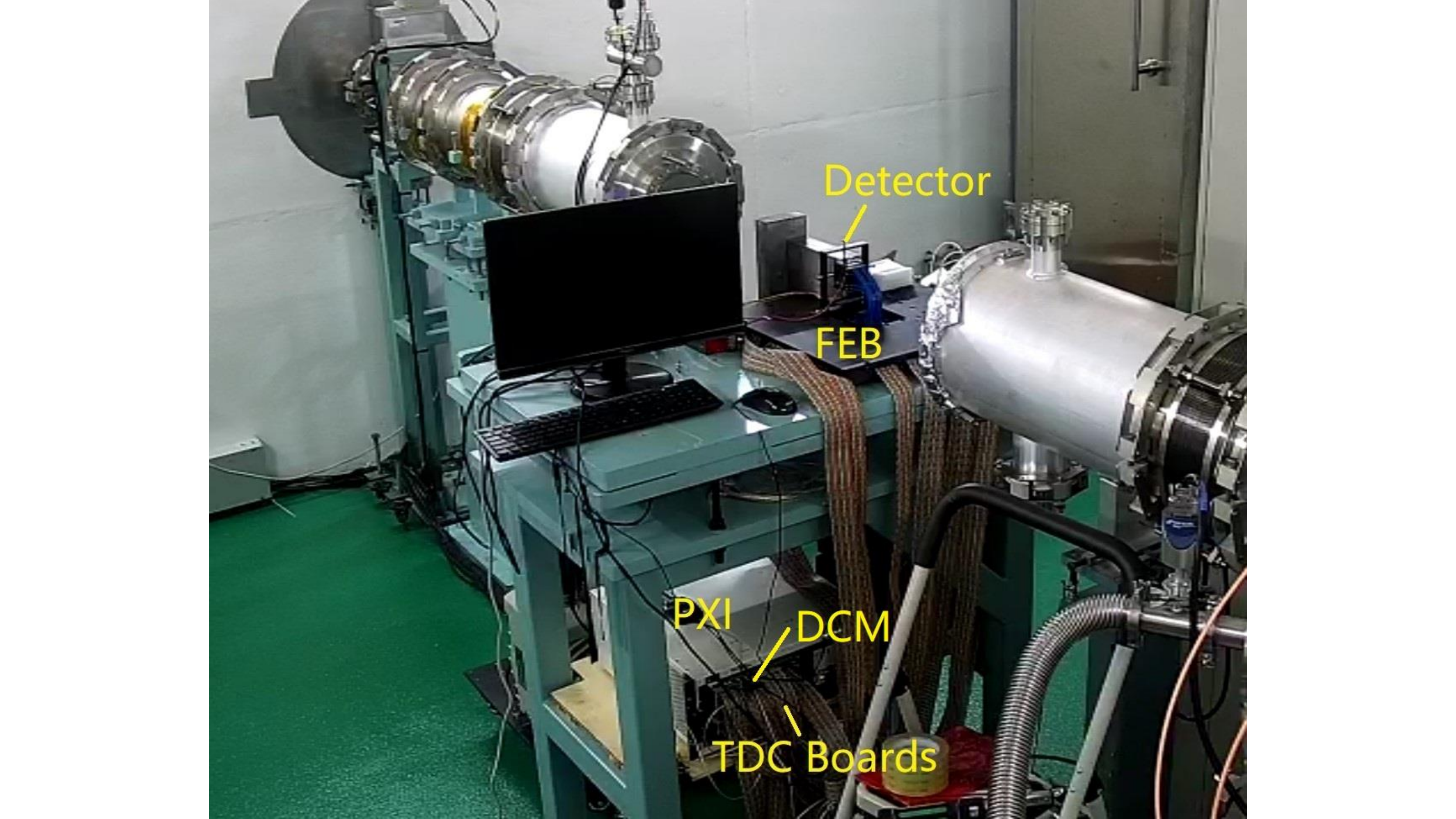}
\caption{Test platform of TOF system in CSNS}
\label{Fig.15}
\end{figure}

We use the system to measure flight time of neutrons in CSNS. Fig.~\ref{Fig.15} shows the test platform of TOF electronics system in China Spallation Neutron Source.

\begin{figure}
\includegraphics[width=3.5in,clip,keepaspectratio]{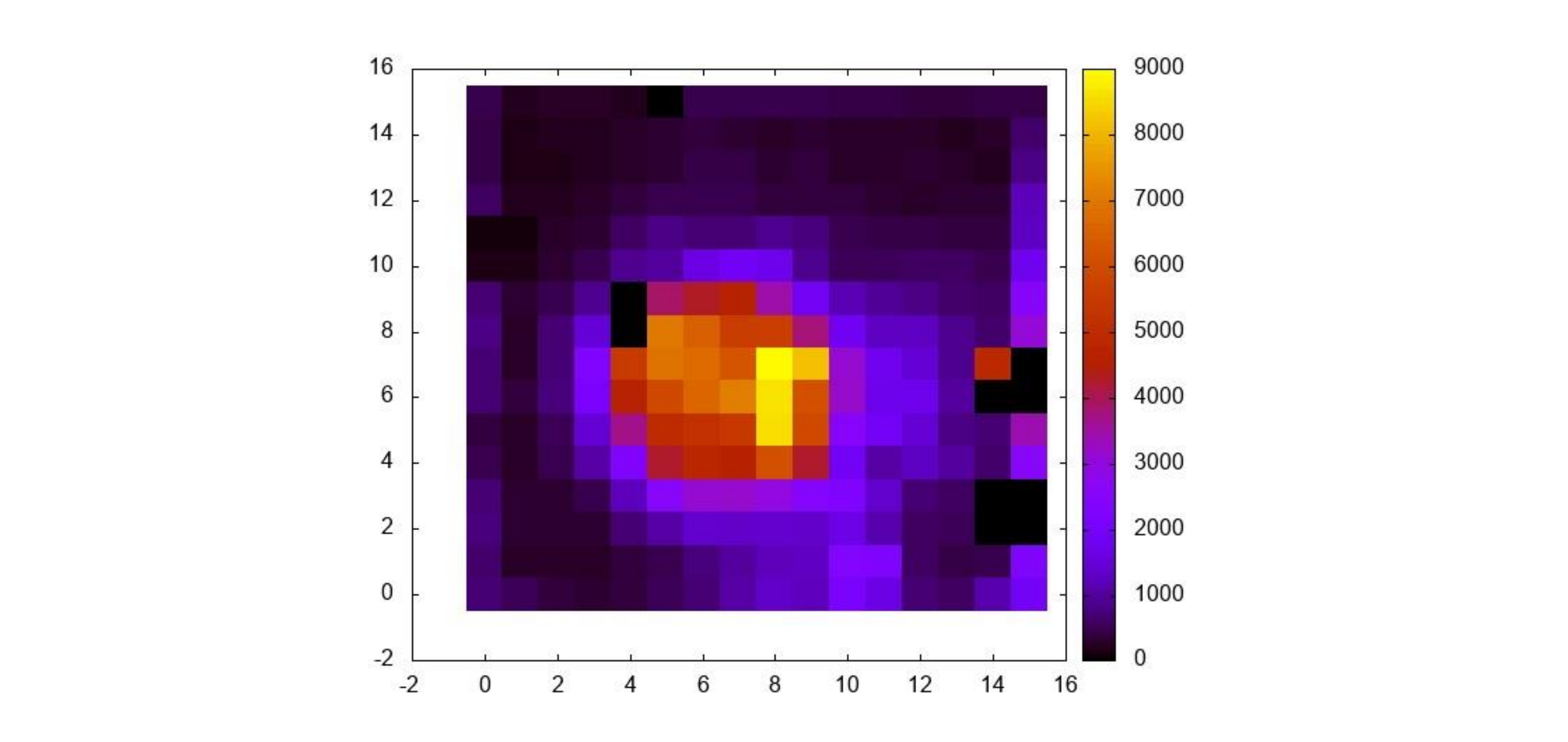}
\caption{The hit map of a circle with 30mm diameter}
\label{Fig.16}
\end{figure}

\begin{figure}
\includegraphics[width=3.5in,clip,keepaspectratio]{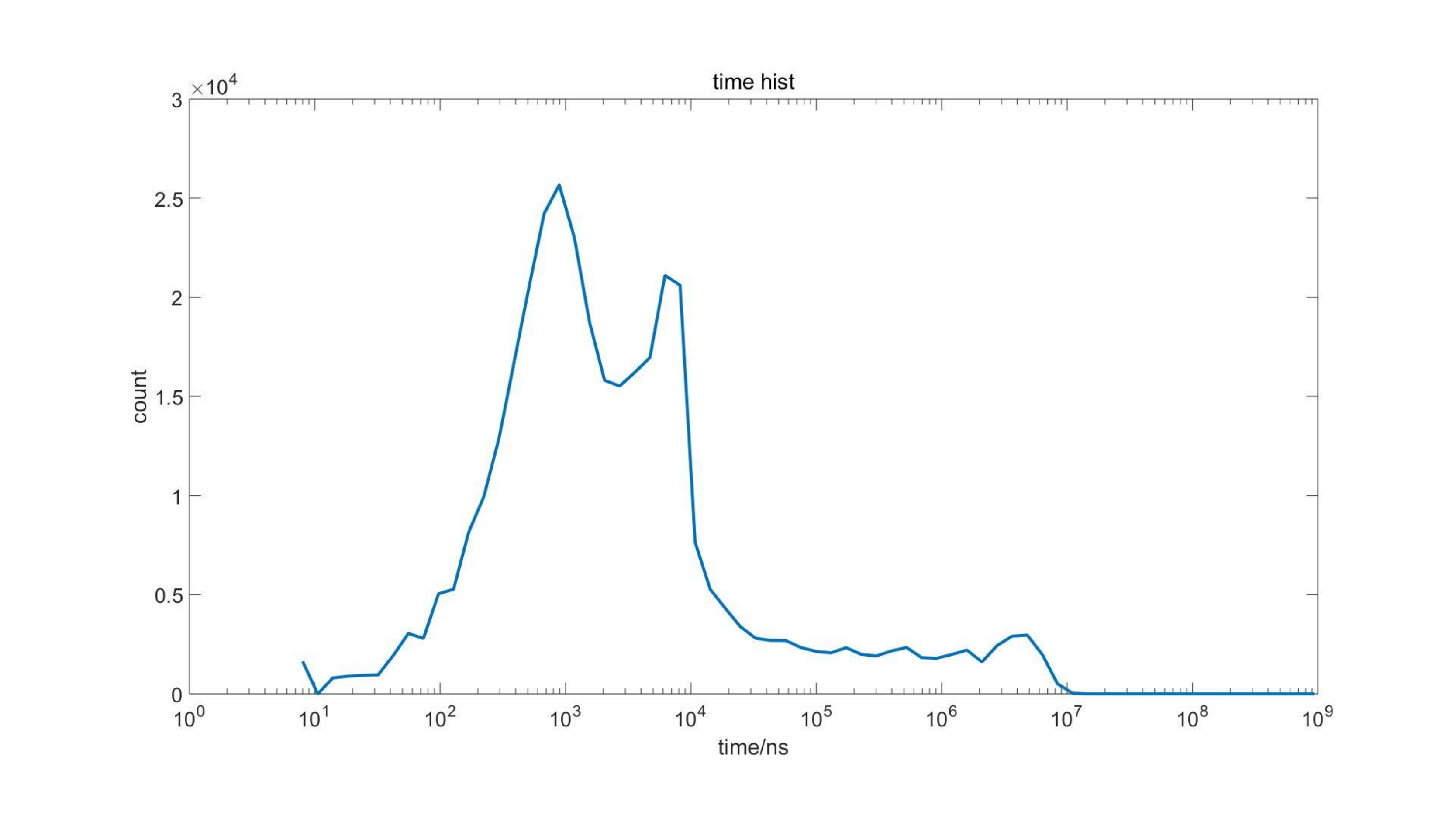}
\caption{The time spectrum of the neutron beam}
\label{Fig.17}
\end{figure}

We got a hit map and a time spectrum. The spot on the detector was controlled to be a circle with 30mm diameter. Fig.~\ref{Fig.16} shows the hit map and Fig.~\ref{Fig.17} shows the time spectrum.

\section{Conclusion}

A 256-channel TOF electronics system has been designed and tested including 256-channel FEB and 64-channel TDC boards. According to the test result, the time measurement can achieve 3.5ns precision and 10ms dynamic range. It can measure the time between the start signal and stop signal well. The system has already been used in experiment of China Spallation Neutron Source. The preliminary test results lay the foundation for the neutron resonance radiography.


%

\section*{Acknowledgment}

The author would like to thank the support from the CSNS Engineering Project and National Key Research Program of China (Grant No.2016YFA0401601).

\ifCLASSOPTIONcaptionsoff
  \newpage
\fi

\end{document}